\documentclass[
    ,final            
  ]
  {aipproc}

\usepackage[OT1]{fontenc}
\layoutstyle{8x11single}
\usepackage{epsfig}
\usepackage{graphicx}
\usepackage{amssymb, amsmath, amsfonts}
\usepackage{epstopdf}
\newcommand{\erf}{{\rm erf}}
\newcommand{\ierf}{{\rm ierf}}

\begin{document}

\title{Force Statistics and Correlations in Dense Granular Packings}

\classification{45.70, 47.50+d, 51.10.+y}
\keywords      {Force and stress correlations, DEM simulations, 
                dense static granular packings, pressure cells}

\author{Stefan Luding}{
  address={Multi Scale Mechanics, TS, CTW, UTwente, 
         P.O.Box 217, 7500 AE Enschede, Netherlands},
altaddress={ e-mail: {\tt s.luding@utwente.nl}}
}

\author{Micha-Klaus M\"uller}{
  address={Multi Scale Mechanics, TS, CTW, UTwente, 
         P.O.Box 217, 7500 AE Enschede, Netherlands}
}

\author{Thorsten P\"oschel}{
 address={Universit\"at Erlangen-N\"urnberg, 
     Institute for Multiscale Simulation, Erlangen, Germany},
altaddress={Cluster of Excellence `Engineering of Advanced Materials', 
     Friedrich-Alexander-University, Erlangen, Germany}
}

\begin{abstract}
  In dense, static, polydisperse granular media under isotropic
  pressure, the probability density and the correlations of particle-wall 
  contact forces are studied. Furthermore, the probability density functions 
  of the populations of pressures measured with different sized circular
  pressure cells is examined.  
  The questions answered are:
  (i) What is the number of contacts that 
  has to be considered so that the measured pressure
  lies within a certain error margin from its expectation value?
  (ii) What is the statistics of the pressure probability density as
  function of the size of the pressure cell?  Astonishing non-random 
  correlations between contact forces are evidenced, which range at least 
  10 to 15 particle diameter. Finally, an experiment is proposed to tackle and 
  better understand this issue.
\end{abstract}

\maketitle


\section{Introduction}

One of the open issues in the field of disordered, random systems like
dense, static granular packings, is the probability density of the
contact forces and their possible long range correlations.  
There is common agreement that the probability for
large forces decays exponentially 
\cite{thornton88,coppersmith96,radjai96b,radjai97,brockbank97,cates98b,radjai98,radjai98b,tsoungui98b,tsoungui99b,silbert01,blair01b,snoeijer03,metzger04,snoeijer05}
but the small forces are much harder
to measure \cite{coppersmith96,liu95,erikson02}, so that there is
still ongoing discussion about the shape of the probability density
for small forces, possible correlations between the forces, and a
predictive model for the force propagation inside dense packings of
frictional particles \cite{tighe08}.

Furthermore, it is observed that the deformation of particle
systems is not affine in general, but displays finite distance 
correlations which are assumed to increase when approaching
the jamming transition 
\cite{cates98b,donev04,corwin05,dauchot05,song05,silbert05,majmudar07,aranson08,zeravcic09,mailman09}. 
However, the issue whether these correlations 
(and possibly anti-correlations due to vortex formation) depend on the 
system size \cite{maloney06} or not \cite{tanguy02,silbert09} is not 
completely resolved yet.

Since the forces that granular particles exert onto their confinement
(walls) strongly fluctuate from one particle to the next, so does also
the local pressure.
When the pressure on the wall is measured with a circular pressure cell,
performing many independent measurements, one obtains a probability 
density of the measured pressure, with its first moment approaching the 
mean pressure, and the standard deviation decreasing with increasing cell size.
While the case of uncorrelated forces is rather behaving as expected,
the case of subtly correlated forces in granular packings leads
to interesting results and probably can be understood with 
advanced statistical methods \cite{jakeman80}, which can e.g. account
for contact number fluctuations.

Thus, a local measurement of the wall pressure can be far away
from the total pressure (or the mean, representative pressure), 
unless ``enough'' particles are contained in the
pressure cell.  Besides the question, how much ``enough'' is, also the
question of the behavior of the width of the pressure probability
density is examined in this study.  The reason to examine the
pressure distribution instead of the force distribution is that the
former is much easier to access experimentally, as will be outlined at
the end of this paper.

In general, more knowledge on the force- and pressure-density
functions is needed for the understanding of pressure measurements
aiming, for example, at a safe design of containers of granular
materials such as silos.

\subsection{Review on force probability}

\begin{figure}[htbp]
  \centerline{\includegraphics[width=8cm,angle=-90]{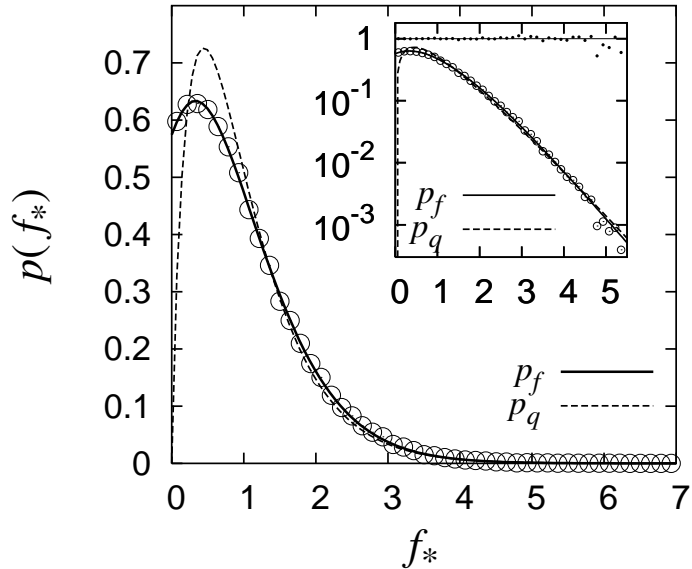}}
  \caption{Normalized probability density $p(f_*)$ plotted against the
    normalized force $f_*=f/\langle f \rangle$. The circles are
    simulation data, and the dashed and solid lines are Eqs.\
    (\protect\ref{eq:qforce}) and (\protect\ref{eq:sforce}), with
    $C=1.8$, respectively. The inset shows the same data in log-scale,
    together with the ``quality factor'', i.e.\ the simulation data
    divided by the fit-function as dots around unity; the fit has
    deviations less than two per-cent for $f_*<3$ and less than ten
    per-cent for $f_*<5$.  The $q$-model, Eq.\
    (\protect\ref{eq:qforce}), is invalid for $f_*<0.2$ and has
    varying deviations of about ten per-cent for $0.2<f_*<5$.  }
  \label{fig:pfsim}
\end{figure}

The simplest model for the force probability density function is the
so-called $q$-model, 
introduced 1995 by Liu, et.\ al.\ \cite{liu95,coppersmith96}, that describes 
the occurrence of force chains in disordered geometric packings of
granular media on the basis of a mean field approximation
\cite{metzger04,snoeijer05}. In a dense
packing of discrete particles, the contact forces that one particle in
a certain layer feels from above plus its weight, determines the sum of
the contact forces on the particles below. The magnitude of the
two forces at the two contacts below, are the fractions $q$ and $1-q$
of the sum.
In general, the weights can also depend on neighboring sites and
contacts can open and close.  The mean field approximation, however,
neglects these dependencies and thus simplifies the model vastly. Eq.\
(\ref{eq:qforce}) gives us the normalized, scaled density function of
the inter-particle forces $f_*=f/\langle f \rangle$, predicted by the
$q$-model:
\begin{equation}
  \label{eq:qforce}
  p_{q}(f_*)=\frac{C^C}{(C-1)!} ~ f_*^{C-1}\exp\left(-Cf_*\right)\,,
\end{equation}
where $C$ is the number of the neighboring particles (below or above).
The weakness of the $q$-model is the improper
prediction of the probability to find small forces, see Fig.\
\ref{fig:pfsim}.

A function, that provides an excellent fit to the simulation data is
\begin{equation}
  \label{eq:sforce}
  p_{f}(f_*)=\left ( 1-a \exp \left ( -\frac{(f_*+b)^2}{8} \right ) \right ) 
                     c \exp\left(-d f_*\right) \, ,
\end{equation}
with the fit-parameters $a=0.983 \pm 0.003$, $b=0.56\pm0.05$,
$c=1.80\pm0.02$, and $d=10.4\pm0.7$. Note that a similar function was
found experimentally, see Refs.\ \cite{mueth98,blair01}. The number of
parameters can be reduced by using the normalization relations $\int p_f df
= 1$ and $\int f \, p_f df=1$
\footnote{the factor 8 in the denominator leads
to some better fit-quality by stretching the Gaussian correction
function}.  The fit-parameters for different densities are given 
in table\ \ref{tab:tabnu}.

\begin{table}
\label{tab:tabnu}
\begin{tabular}{l|l|l|l|l}
\hline
$\nu$ & $a$ & $b$ & $c$ & $d$ \\
\hline
0.80 & 0.995298 & 0.477783 & 14.089 & 1.91749\\
\hline
0.70 & 0.971913 & 0.68069 & 7.72909 & 1.68119 \\
\hline
0.68 & 1.00345 & 1.05915 & 5.127 & 1.54689 \\
\hline
0.66 & 0.970768 & 1.11105 & 4.25643 & 1.47278\\
\hline
0.64 & 0.902126 & 0.507015 & 6.3781 & 1.57989\\
\hline
$e$ [\%] & $1-4$ & 15-23 & 11-20 & 2-4 \\
\hline
\end{tabular}
\caption{Fit results for different volume fractions $\nu$, where
the row $e$ contains the typical relative error of the 
respective fit-parameter.}
\end{table}

\section{Theory}
\label{sec:theory}

\begin{figure}[htbp]
  \centerline{\includegraphics[width=6cm]{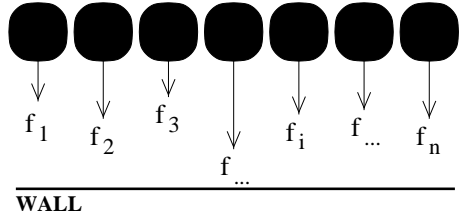}}
  \caption{The forces of particles that touch the walls fluctuate
    around a mean value.}
  \label{fig:aequiv}
\end{figure}

Let us consider a cubical container filled with granular material
under isotropic, hydrostatic pressure. Each particle $i$ of the $N_w$
particles that touch a wall exerts a force $f_{i}$ (see Fig.\
\ref{fig:aequiv}) onto the wall and contributes to a (finite: $1 \le i
\le N_w$) population of forces $p(f)$ with mean $\mu_w=\left<f\right>$
and standard deviation $\sigma_w$, where the subscript $w$ refers to
the population of {\em all} wall-particle forces.  If we select one
sample $j$ of size $n$ out of this population by applying a circular
sensitive area $D(R)=\pi R^{2}$ that includes the $n$ forces that are
acting on this area, we obtain the pressure
\begin{equation}
P_j(R)=P_j(n_j(R))
 = \frac{ (1/n) \,\sum_{j=1}^{n} f_{j} }{ D(R)/n } 
 = (1/n) \sum_{j=1}^{n} f_{j}/D_1 ~, 
\end{equation}
with the area per particle, $D_1=D(R)/n$, and $n=n_j(R)$. 
Note that such areas have to be selected such that their center
-- on the selected wall -- is at least a distance $R$ away from any
other wall. Taking many samples, $m \gg 1$, will result in a population
of pressure values with probability density $p_R(P)$ around the
mean $\mu_P=\left<P_j(R)\right>=(1/m)\sum_{j=1}^{m} P_{j}(R)$, with
standard deviation
$$\sigma_P = \sqrt{ \left<P_j^2(R)\right> - \mu_P^2 } ~,$$ 
where the subscript $P$ refers to the population of pressures. 

\begin{figure}[htbp] 
  \centerline{\includegraphics[width=6cm]{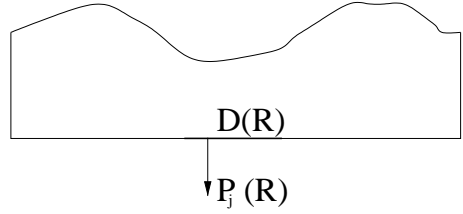}}
  \caption{The average force of sample $j$, divided by the sensitive area 
    $D(R)/n$ per particle, of this detector, leads to a pressure 
    $P_{j}(R)$ corresponding to one measurement $j$.}
  \label{fig:box}
\end{figure}

\subsection{Central Limit Theorem}

According to the central limit theorem (CLT), and for the
corresponding assumptions, the probability density function $p_R(P)$
of our population of samples $P_{j}(R)$ provides the same expectation
value as for the original population $\mu_P \approx \sum_{i=1}^{N_w}
f_{i}/A$, where $A$ denotes the
total surface of the confinement and the sum goes over all particles
in the system touching the walls 
\footnote{This is valid under the assumption that all
  wall-particles are similar, i.e.\ there is no inhomogeneity in the
  forces on the wall particles, e.g.\ as function of distance from
  another wall in the edges of the cuboid.}.
The CLT also tells us that the
probability density function becomes more and more Gaussian:
\begin{equation}
  g_R\left(P\right)=\frac{1}{\sqrt{2\pi} \sigma_{P}}
  \exp \left [ -\frac{1}{2} \left ( \frac{P-\mu_P}
                                         {\sigma_{P}}
                            \right )^{2} \right ] ~,
\label{eq:gRP}
\end{equation}
the larger we chose $n$, i.e., by increasing the detector size $R$. 
The standard deviation then equals $\sigma_{P}=\sigma_{P}(R)=(n(R) \,
\sigma_{w})/(D(R)\sqrt{n(R)}) =\sigma_w/(D_1
\sqrt{n(R)})=:P_f/\sqrt{n}$, with the pressure $P_f$ that corresponds
to the standard deviation of the force density function scaled by the
area of the pressure cell per particle.


\subsection{Confidence Intervals}

Integration from $-\infty$ to $+\infty$ of the normalized
distributions $p_R(P)$ and $g_R(P)$ gives unity.  Now, to gain more
advanced statistical predictions about the pressure distribution, let
us consider a lower and an upper integration limit
$z_{\alpha/2}<\mu_P$ and $z_{1-\alpha/2}>\mu_P$, respectively,
such that the integral over Eq.\ (\ref{eq:gRP}) equals 
\begin{equation}
\label{eq:erf1}
1-\alpha 
 = \int \limits_{z_{\alpha/2}}^{z_{1-\alpha/2}} g_R(P)~dP              
 = 1-\int \limits_{-\infty}^{z_{\alpha/2}} g_R(P)~dP
      -\int \limits_{z_{1-\alpha/2}}^{\infty} g_R(P)~dP         
 = \frac{1}{2}\left[ \erf\left( \frac{z_{1-\alpha/2}-\mu_P}
                                       {\sqrt{2} \sigma_{P}} 
                            \right)
                      +\erf\left( \frac{\mu_P-z_{\alpha/2}}
                                       {\sqrt{2} \sigma_{P}} 
                            \right) 
                \right] \,. 
\end{equation}
If the integration limits are chosen such that a fraction $\alpha/2$
lies outside of the integration range, both to the left and the right,
the integration limits correspond to the confidence interval $2
\delta_\alpha=z_{1-\alpha/2}-z_{\alpha/2}$; a fraction $1-\alpha$ of
the $n$ measurements $P_j$ lies within the confidence interval.

Keeping $n=const.$ and considering $\alpha=0$ (probability for finding
the measured value in-between our limits then equals 1), we expect
that the interval of confidence tends to infinity.  On the other hand,
for $\alpha=1$ (probability vanishes), one expects $\delta_{\alpha}
\rightarrow 0$.

Due to the symmetry of the Gaussian distribution, one can compute
\begin{equation}
\delta_{\alpha}:=z_{1-\alpha/2}-\mu_P=\mu_P-z_{\alpha/2} 
\end{equation}
explicitly, using the relation:
$\ierf(1-\alpha) = \delta_\alpha /(\sqrt{2} \sigma_P) = \delta_\alpha \sqrt{n} /(\sqrt{2} P_f)$
such that:
\begin{equation}
  n=2 \left ( {P_{f}}/{\delta_{\alpha}} \right )^{2} ~ \ierf^{2} \big(1-\alpha \big) ~,
  \label{eq:n-delta}
\end{equation}
where the pressure $P_f=n(R) \sigma_w/D(R)=\sigma_w/D_{1}$ is, of
course, a constant for given geometry and $p(f)$.  The general
equation (\ref{eq:n-delta}) gives us information about how many
particles $n$ need to be considered in a measuring process, so that 
a measurement $P_{j}(R)$ lies in-between the error margin 
$\pm \delta_{\alpha}$ with probability $1-\alpha$. Note, that
$\delta_{\alpha}$ and $n$ depend on the size $R$ of the
sensitive area $\delta_\alpha \propto 1/\sqrt{n(R)} \propto 1/R$.

\subsection{Explicit predictions}

\subsubsection{$q$-model}

Now we use the distribution predicted by the $q$-model for the
analysis. The first two moments obtained from Eq.\ (\ref{eq:qforce})
are $\overline{f}=\langle f \rangle$ and $\overline{f^2}=\frac{C+1}{C}
\langle f \rangle^2$, which leads to the standard deviation
$p_q(f)$, namely $\sigma^q_{f}=\left (
  \overline{f^{2}}-\overline{f}^{\,2} \right )^{1/2} =\sqrt{1/C}
\,\langle f \rangle$.  Considering Eq.\ (\ref{eq:qforce}) as a finite
population and applying the CLT by taking many samples, one can then
replace $P_f^2$ in Eq.\ (\ref{eq:n-delta}) to get
\begin{equation}
  n^{(q)}  =  \frac{2}{C} \left ( \frac{\langle f \rangle}{D(R)/n} \right )^2
          \ierf^{2}\big(1-\alpha \big) \frac{1}{\delta_{\alpha}^{2}} 
           =  (2/C) \, \ierf^2(1-\alpha) ( P_f / \delta_\alpha )^2 ~.
  \label{eq:n-delta2}
\end{equation} 

\subsubsection{Best Fit}

Now we use the distribution as obtained by our fit to the data for the
analysis.  The first two moments obtained from Eq.\ (\ref{eq:sforce})
are $\overline{f} \approx \langle f \rangle$, $\overline{f^2}=1.61
\langle f \rangle^2$ and the corresponding standard deviation can be
computed to $\sigma^{fit}_{f}=0.78 \langle f \rangle$. In comparison
with these results the $q$-model provides $\sigma^q_{f}=0.76 \langle f
\rangle$ for a dense granular packing with $C=1.8$ corresponding to a
special geometry. 

\section{Simulation}

\label{sec:simulation}
\begin{figure}[htb]
  \centerline{\includegraphics[width=10cm,clip]{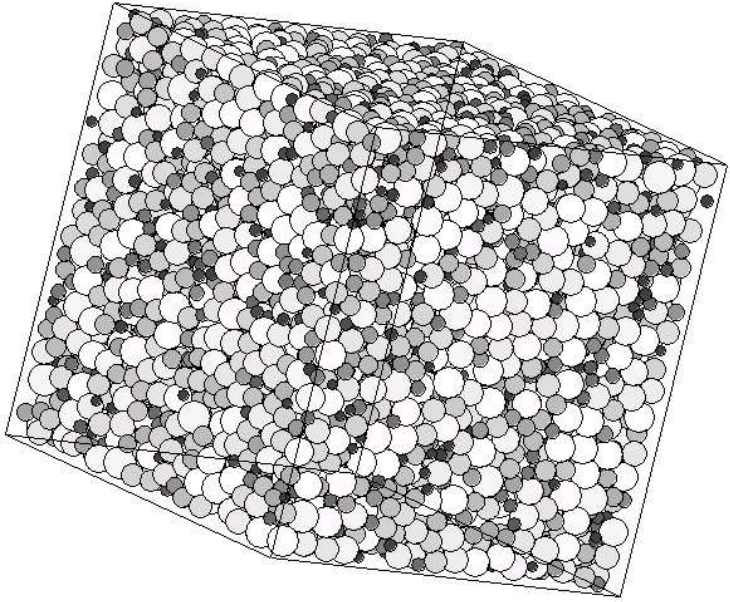}}
  \caption{Snapshot of a dense, polydisperse assembly of $N=8000$
    particles confined in a cuboid. The volume fraction here is $\nu
    \approx 0.7$ and the particle sizes are grey scaled (bright
    particles are big, dark particles are small).}
  \label{fig:dense}
\end{figure}

\begin{figure}[htb]
  \includegraphics[width=6cm,clip]{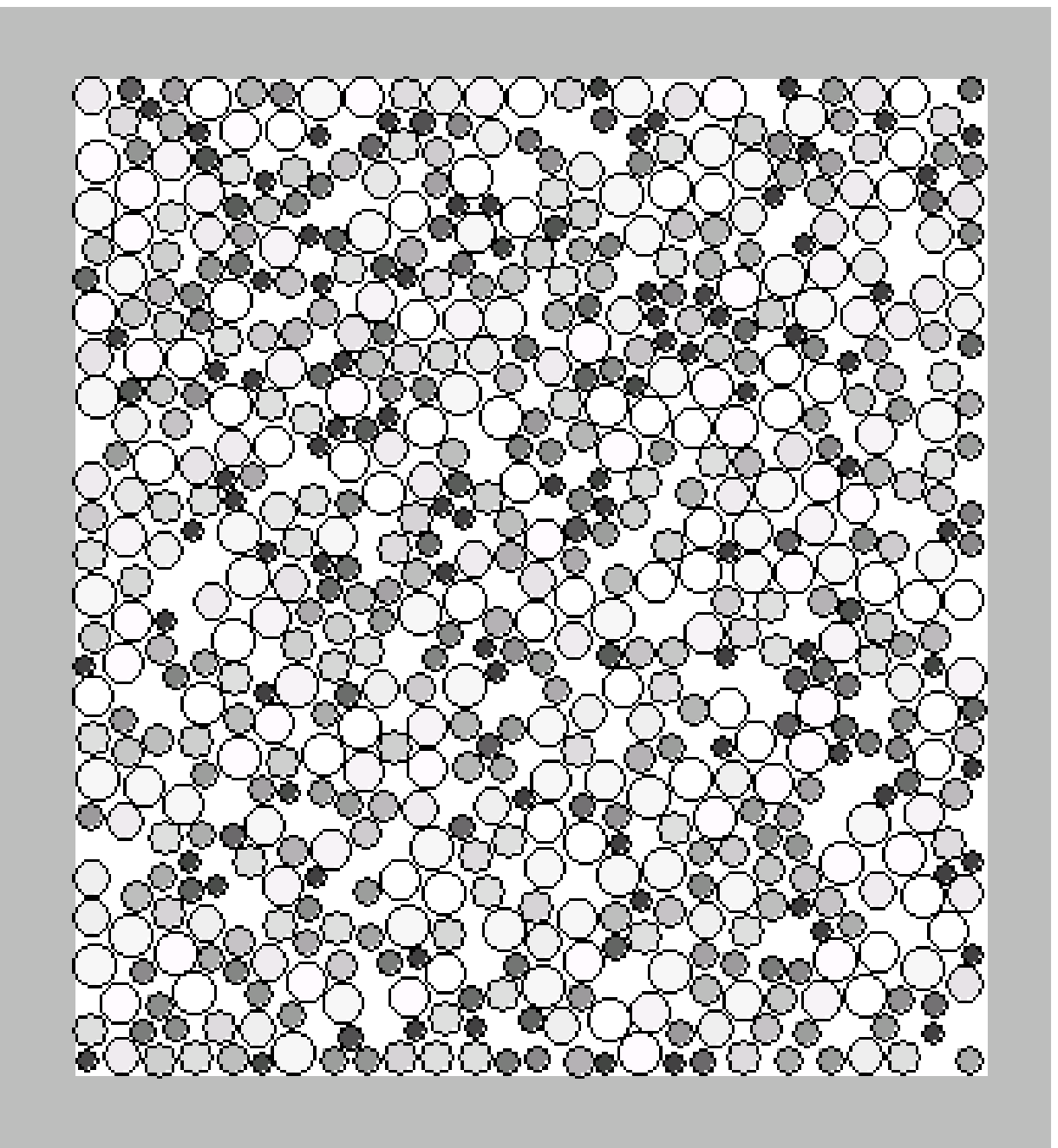}
  \includegraphics[width=6cm,clip]{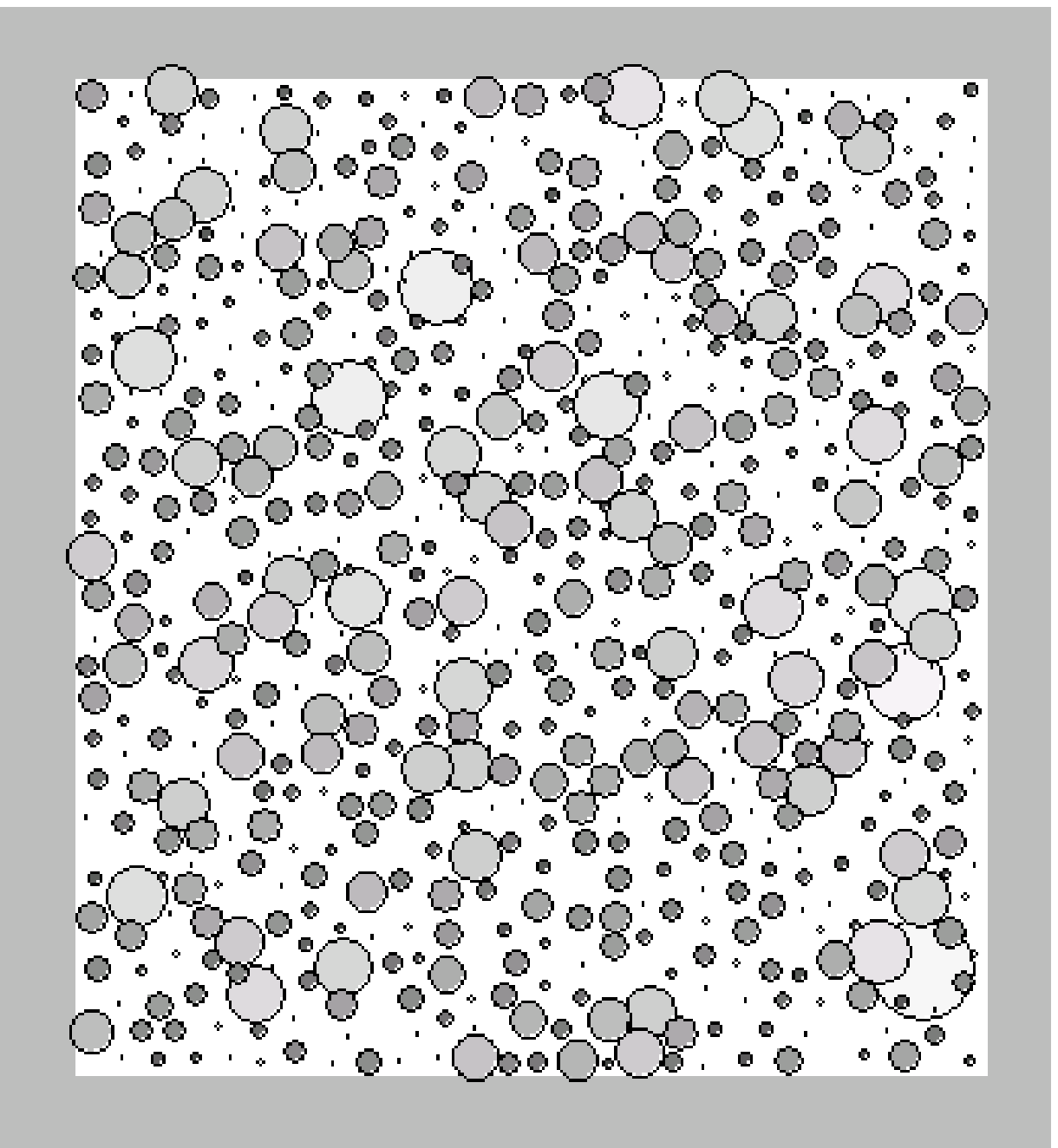}
  \caption{View on one of the walls for a packing with $N=20000$
    particles with $\nu \approx 0.67$.  (Left) Each circle is a
    particle in contact with the wall; the color coding is the same as
    in Fig.\ \protect\ref{fig:dense}.  (Right) Same data as (left),
    but here each circle radius is scaled with the force exerted by
    the particle on the wall. (Big, bright circles correspond to large
    forces, whereas small, dark circles correspond to small forces on
    the wall).}
  \label{fig:p-on-wall-y1}
\end{figure}

\subsection{The System}

The systems studied in the following contain $N$ perfectly 
spherical particles with radii $r_{i}$ drawn from a 
homogeneous distribution $r_{i} \in [r_{\rm min},r_{\rm max}]$, 
where we used $r_{\rm max}/r_{\rm min}=2$ and $3$, and the mass 
$(4/3)\pi r_{i}^{3}\rho$, where $\rho$ represents
the uniform material density of the particles.  Since the mass is not
relevant in the static limit, we refer to density only with the
dimensionless volume fraction
$\nu_{f}=\frac{4}{3}\pi\sum_{i}^{N}r_{i}^{3}/\prod_{\beta}^{3}L_{\beta}$,
where $L_{\beta}$ denotes the length of the simulation volume in
direction $\beta \in (x,y,z)$. All of the simulations were done in a
cuboid volume which is limited by walls that repell touching
particles. Fig.\ \ref{fig:dense} gives an example of a typical static
and dense granular sample, while Fig.\ \ref{fig:p-on-wall-y1}
shows a representative subset of particles that touch 
a wall (Left) and the corresponding magnitudes of forces (Right),
quantified by the radius of the circles.

\subsection{Molecular Dynamics}

The simulations were performed by means of a molecular dynamics (MD)
code in three dimensions, without tangential forces like friction. MD
simulations are characterized by discretizing time into timesteps
$\Delta t$ and solving the $Newton$'s equations of motion for each
particle.  In each integration step, the new position of the particles
is computed from its previous and present positions and accelerations
due to forces currently acting on it \cite{allen87}.  MD is also
referred to as discrete element method (DEM) or as soft-sphere model, 
i.e.\ the repulsive forces $f^{n}$ normal to the plane through the 
point of contact depend on the overlaps of the spheres, that replace 
the contact deformation. 
The linear spring-dashpot (LSD), the non-linear $Hertzian$ model and a
hysteretic force model can be used \cite{luding98c},
among many others.  For particle-wall
contacts, the same spring constant were used as for contacts between
particles.  Here, only the LSD results are discussed because no
qualitative differences could be evidenced for the different contact
laws.

\subsection{Initial Configurations and Relaxation}

As initial configurations, $N=8000$ or $N=20000$ poly-disperse
particles with random initial velocities were placed on a regular
cubic lattice with low total density.  Due to the free space between
the particles, the initial order is forgotten and the particles
disspate energy during collisions. Eventually, the dense, relaxed and
disordered granular packing is obtained by either applying hydrostatic
pressure on the walls or by growing the particles.  Due to the
dissipative nature of the contact law, energy is dissipated and the
system reaches a static configuration, where we use as criterion the
ratio of kinetic and potential energy, $E_{\rm kin}/E_{\rm
  pot}=\epsilon$, with $\epsilon \le 10^{-7}$.  If $\epsilon$ is
small, the particles are typically almost at rest and the major
contribution to the total energy stems from the contact potential
energy between particles and between particles and the walls.  Since
the results were identical for the two preparation procedures, we only
mention that there are more alternative ways of achieving a static
packing, see \cite{bagi03}.

\section{Results}
\label{sec:results}

The force- and pressure density functions are gained from the
(repulsive) forces between all particles that touch the walls.  Note
that the scaled force-density function for all particles in the bulk
was identical to the force-density for particle-wall forces, within
the considerably larger fluctuations of the latter, only the mean
particle-particle force is smaller than the mean particle-wall force.
Since we are interested in the experimentally accessible pressure
measured at the wall, we will not present bulk data in the following.

When a circular measuring area with radius $R$ is put around each
wall-particle (for all wall particles with distance larger than
$R_{\rm max}$ from any other wall), one obtains a set of $P_{j}(R)$
values and, from these, can compute the mean values and standard
deviations.  A typical set of wall-particles is shown in Fig.\
\ref{fig:p-on-wall-y1}.  Note that $R_{\rm max}$ is introduced, so
that the same set of wall-particles is used for the computation
independent of $R$ and, if $R_{\rm max}/r_{\rm max}$ becomes larger
than 12-13, the statistics becomes bad, since too few particles close
to the center of the wall are taken into account as the centers of
pressure cells.

\begin{figure}[htb]
  \centerline{\includegraphics[width=6cm,angle=-90,clip]{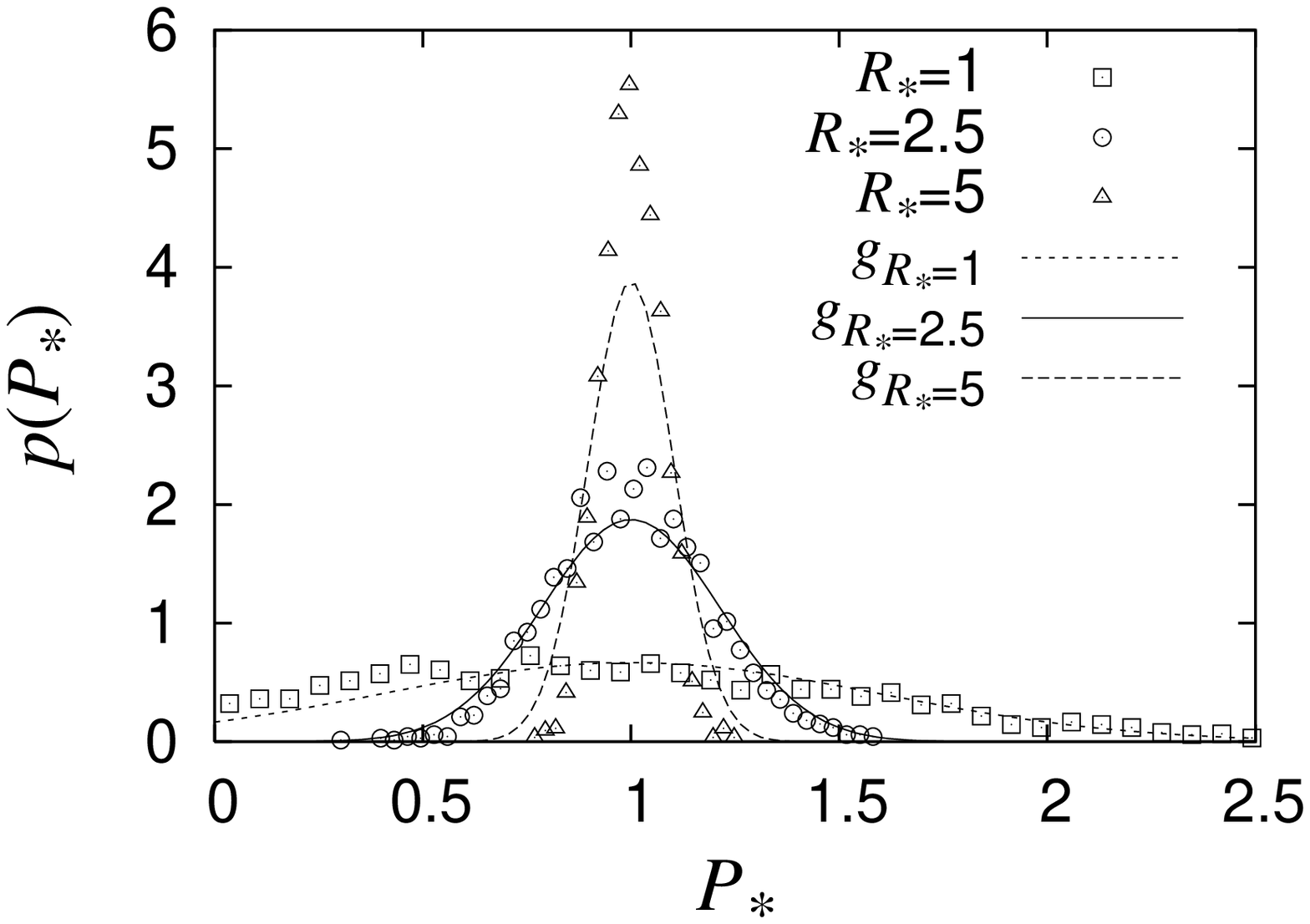}
\includegraphics[width=6cm,angle=-90,clip]{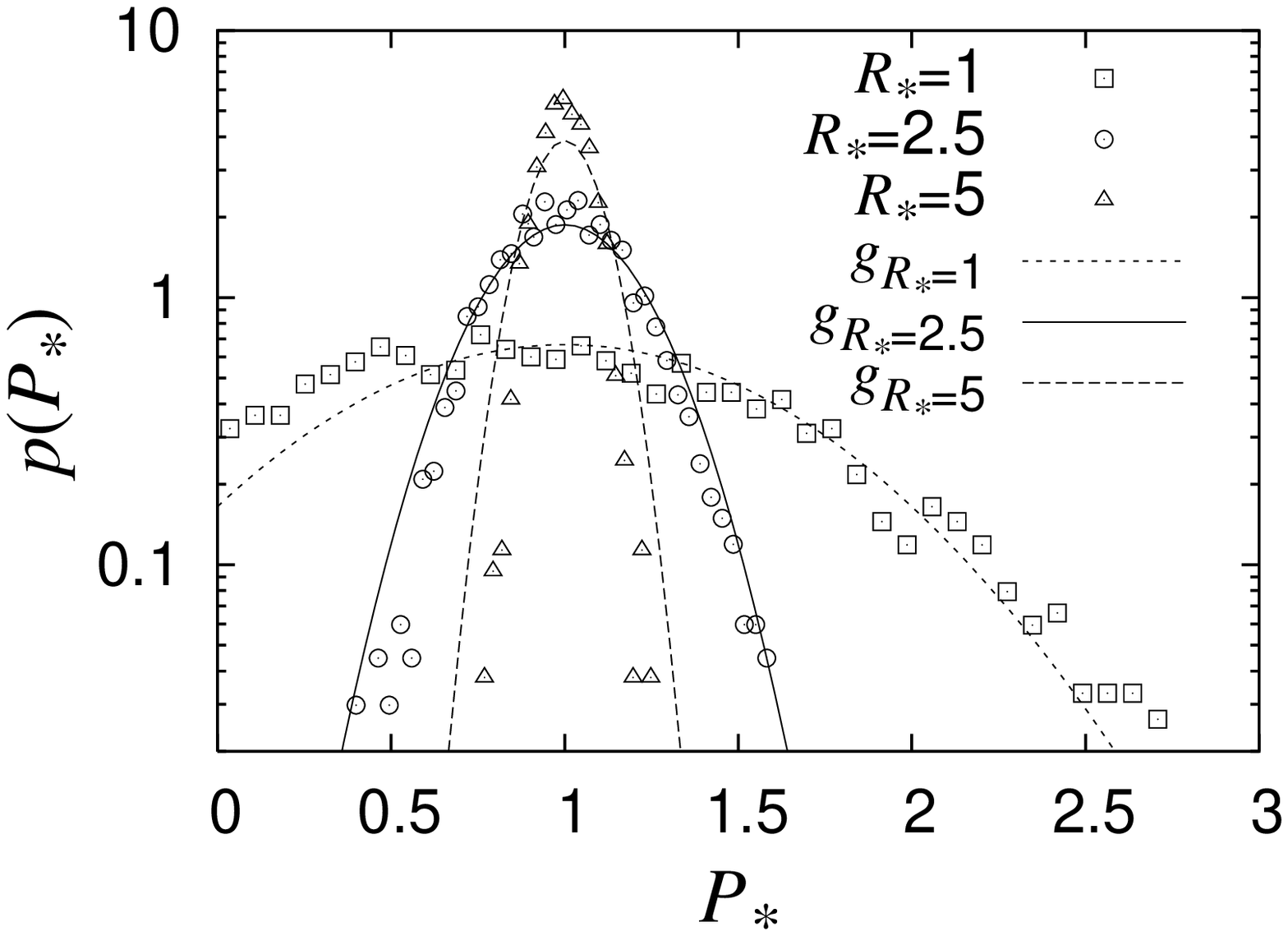}}
\caption{Probability density for the pressures from many samples with
  sizes $R_*=R/r_{\rm max}=1.0$, $2.5$, and $5.0$. Points are
  simulation data and lines correspond to Eq.\ (\protect\ref{eq:gRP})
  The left panel shows the same data as the right, only the latter
  has a logarithmic vertical axis.
}
  \label{fig:ZGWS}
\end{figure}

\begin{figure}[htb]
  \centerline{\includegraphics[width=6cm,angle=-90,clip]{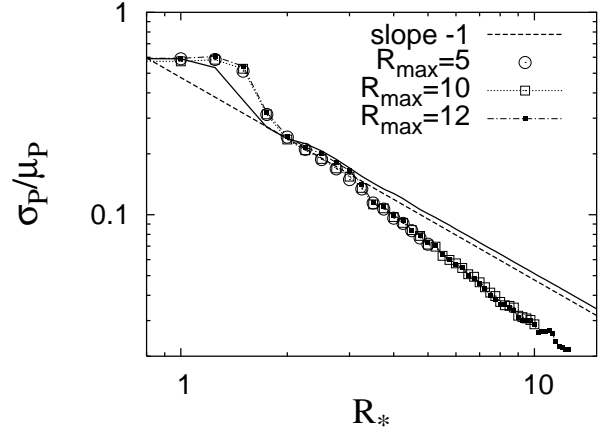}}
  \caption{Scaled standard deviation $\sigma_{P}/\mu_P$ plotted
    against the size of the of cells $R_*$ for different $R_{\rm
      max}$.  The symbols are simulation data, while the solid and the
    dashed lines correspond to $\sigma_{P}/\mu_P = \sqrt{n} \sigma_w
    /(\pi R^2 \mu_P)$ and $\sigma_{P}/\mu_P \propto 1/R_*$,
    respectively.  }
  \label{fig:sigma-Rb}
\end{figure}

For each sample with given $R$, from the corresponding pressures
$P_{j}(R)$, the histograms are obtained, as shown in Fig.\
\ref{fig:ZGWS} for three different $R$-values.
The larger the cell size, the more the pressure density function
appears Gaussian and, as Fig.\ \ref{fig:sigma-Rb} shows, the smaller
the standard deviation $\sigma_{P}$ becomes.  Note, however, the
interesting fact, that the decay of $\sigma_{P}$ is steeper than the
simple relation $\sigma_{P} \propto 1/R$, expected from the central
limit theorem.

\begin{figure}[htb]
  \centerline{\includegraphics[width=6cm,angle=-90,clip]{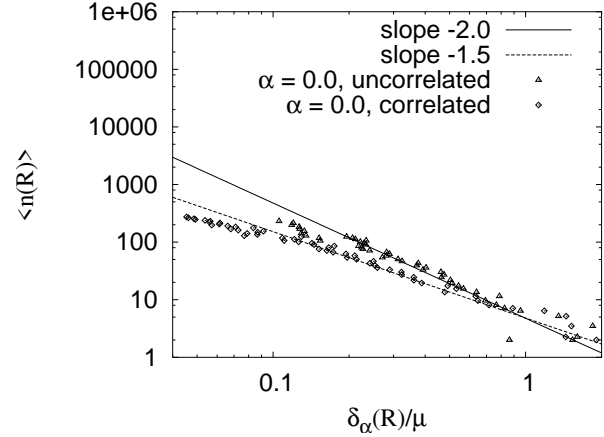}}
  \caption{Number of particles averaged over many samples with given
    radius $R$, for many radii drawn from the interval of confidence
    for $\alpha=0$, scaled by the population mean. The simulations were
    performed with $N=20000$ particles and LSD. The data with slope of
    $-1.5$ gained by our MD simulations reveal correlations between
    the forces of the wall-particles because they do not match to
    theory, see Eq.\ (\ref{eq:n-delta}), that predicts a slope of
    $-2$. Data with slope of $-2$ are ``faked'' by replacing the forces
    by uniformly distributed random forces.}
  \label{fig:uncorr}
\end{figure}

Fig.\ \ref{fig:sigma-Rb} also contains steps for small cell sizes due to 
integer jumps in the particle number $n(R)$. For larger $R$, the change of
$\sigma_P$ becomes smooth and independent of $R_*$.
For every kind of distribution of the population we should get --
according to CLT -- a Gaussian distribution of the $P_{j}(R)$ 
around the population mean. Eq.\ (\ref{eq:n-delta}) is a consequence of 
these simplifications. But our simulations show $\left<n\right> \propto
\delta_{\alpha}^{-1.5}$ (see Fig.\ \ref{fig:uncorr}) or, what is
aequivalent, deviations from the Gaussian shape of the curves for
large $R$ as well. Note, that for very small radii only a very small
amount of particles will contribute to the pressure, i.e., at least, 
the central particle will contribute to the pressure. 


These correlations vanish if we replace the forces in our data files 
by randomly and uniformly distributed forces.  
These (faked) data then show agreement with
Eq.\ (\ref{eq:n-delta}), see also Fig.\ \ref{fig:uncorr}, 
as expected, and finally confirm that correlations
do occur in our systems (samples of particles that 
contribute to the pressure measured at the walls). 

\subsection{Correlation Function}

An alternative way to examine the correlations in a set of forces is
to compute the correlation function
\begin{equation}
C(r) = \frac{\langle f_i f_j(r) \rangle}{\langle f_i \rangle \langle f_j(r)\rangle} ~,
\end{equation}
where the $f_\alpha$ are the forces of the center particles, for
$\alpha=i$, and of all other particles at distance $r$ in the pressure
cells, for $\alpha=j$.
\begin{figure}[htb]
  \centerline{\includegraphics[width=6cm,angle=-90,clip]{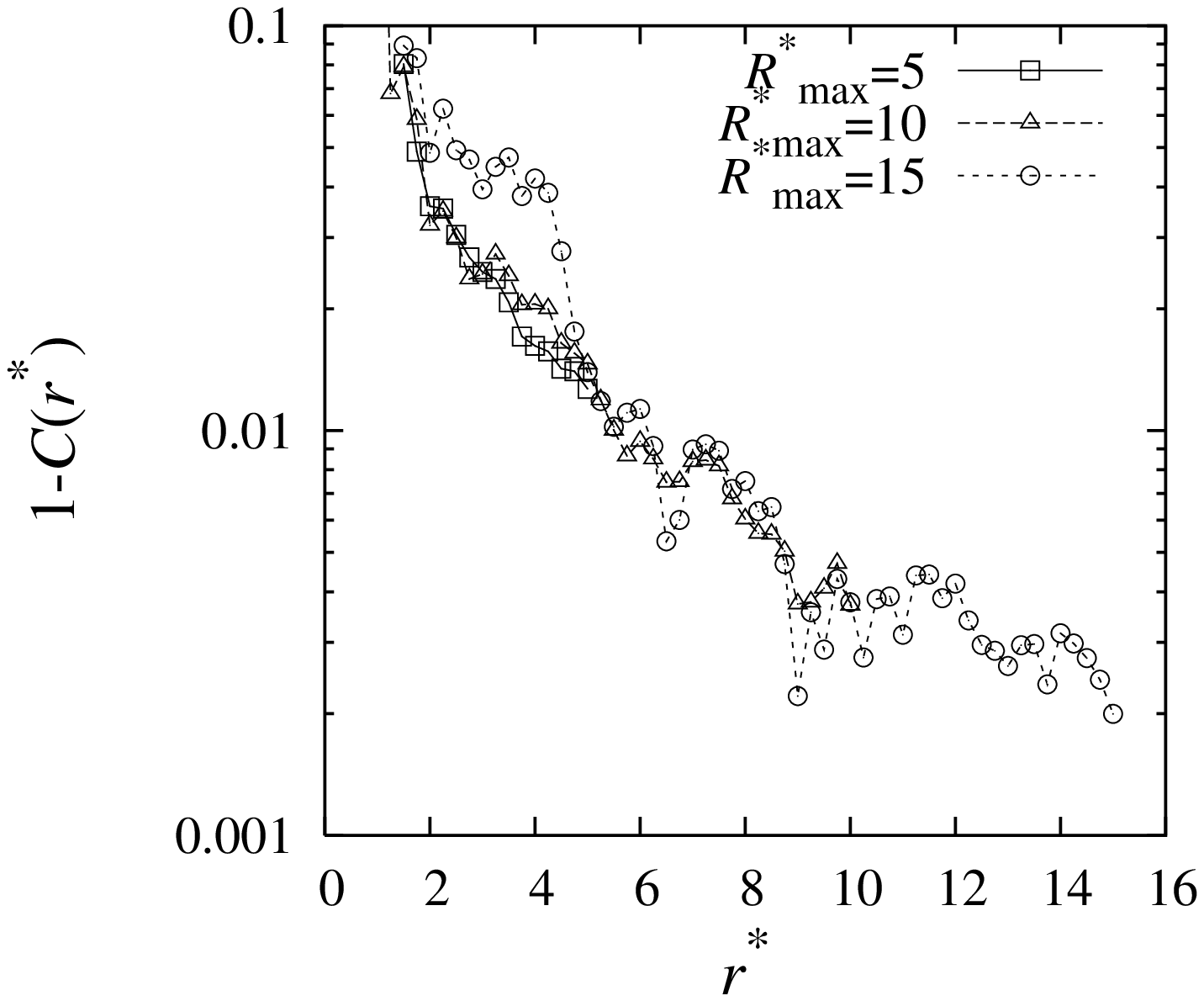}
  \includegraphics[width=6cm,angle=-90,clip]{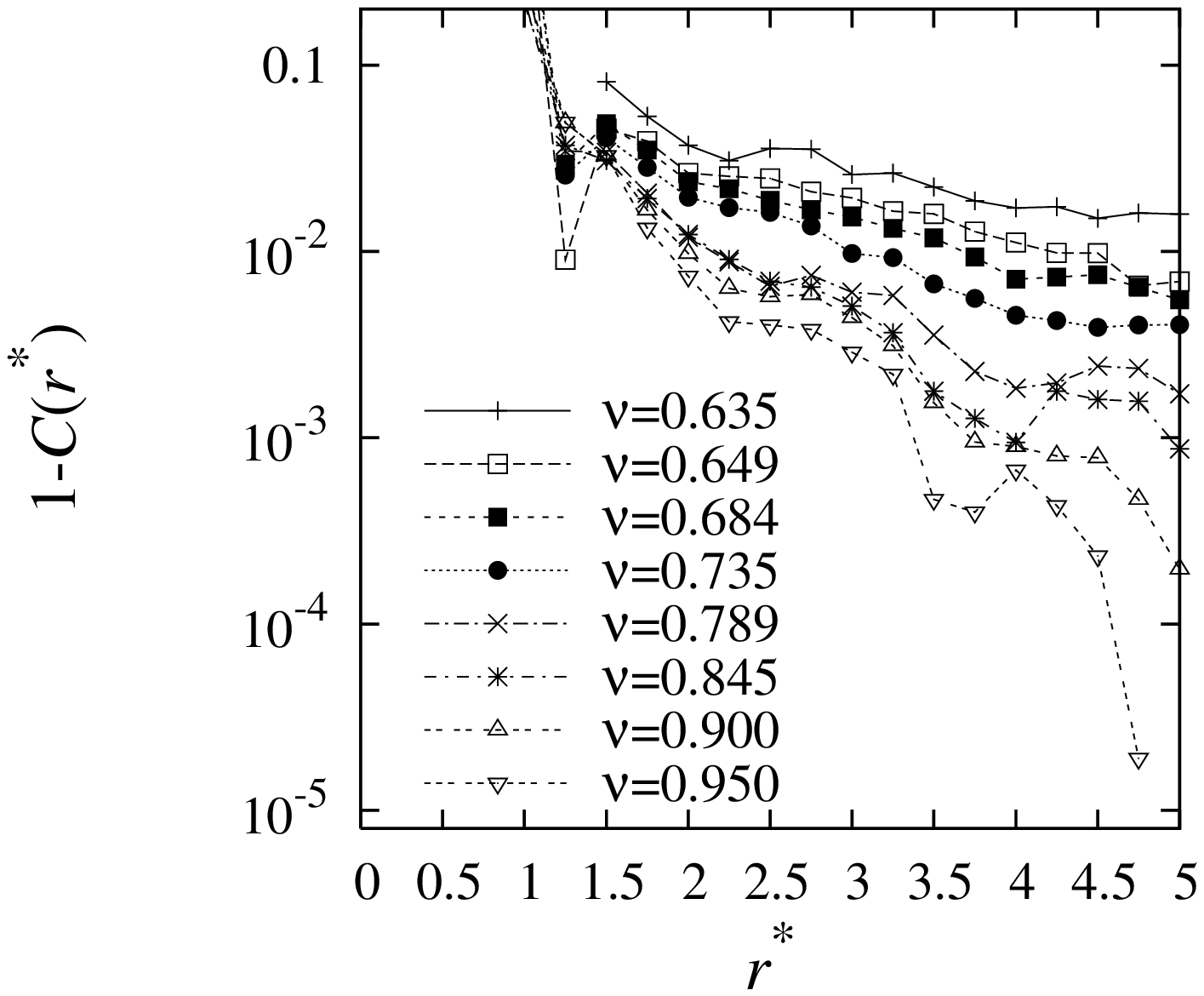}}
\caption{(Right) Correlation function $1-C(r^*)$ as function of the
  scaled distance from the center of the pressure cell, $r^*=r/r_{\rm
  max}$, for different sizes of the pressure cells $R^*_{\rm max}$, as 
  given in the inset. (Right) Correlation function $1-C(r^*)$ for 
  different densities $\nu$, as given in the inset.}
  \label{fig:corrf}
\end{figure}
%
%
As displayed in Fig.\ \ref{fig:corrf} (Left), the correlations decay
about two orders of magnitude within a distance of approximately 15
particle (maximal) radii.  For small $R^*_{\rm max}$, this contrasts 
the results from the confidence interval (or standard deviation) 
measurements, where no change of behavior could be evidenced.
These results are not changing for larger $R^*_{\rm max}$.

The data for different densities in Fig.\ \ref{fig:corrf} (Right)
show that the correlations decay much faster for larger densities.
Note that there is no qualitative difference visible when one examines
$\sigma/\mu_P$ as function of $R^*$ for different densities: The
absolute values decrease with increasing density, but the slope of
$-3/2$ remains independent of the density (data not shown).
Thus, in conclusion, the pressure cell approach is
capable of detecting a different type of correlations, 
which is not caught by the classical correlation function
approach.

\section{Conclusions}
\label{sec:conclusions}

Our main objective was achieved by determining how the number of
particles contributing to the pressure on a given sensor correlate
with the interval of confidence, which is a measure for the width of
the pressure density functions. As expected, the confidence interval 
increases the smaller the detector size is chosen, i.e., 
the worse the statistics becomes. 
Surprisingly, the relation
$\left<n(R)\right>\propto\delta_{\alpha}(R)^{-1.5}$ is observed, 
i.e.\ for a certain amount of particles, given a desired probability 
$1-\alpha$, the measured pressure values can be expected to be closer 
to the average, $\left<P(R)\right>$, as assumed from CLT.
Thus, our simulations predict a better confidence in 
measured data. 

As possible reason for this, one has the  correlations between 
forces exerting by close-by the particles on the walls. 
These correlations range over a rather long distance. The rather
limited sensor size (and the related fluctuations of particle numbers)
could excluded as source of this effect, 
because we found $\left<n(R)\right>\propto\delta_{\alpha}(R)^{-2}$
for a fully random (uncorrelated), uniform force distribution.

Finally, we note that our systems were rather small, so that we 
cannot exclude the possibility that our observations are due
to a finite size effect. Therefore, much larger simulations 
should be performed to confirm the present results and 
better understand the source of the correlations as evident 
from the width of the pressure probability density function.
Furthermore, we analysed our simulation data only at the walls,
so that a direct experimental access to the same information is
possible. In addition, the pressure correlations presented here
should be confronted to bulk stress data in order to learn if
this is a wall effect or intrinsic also to bulk particulate systems
far away from the walls.


\begin{theacknowledgments}
Helpful discussions with K. Hopcraft and N. P. Kruyt
are gratefully acknowledged.
This work was made possible by the financial support of
the Deutsche Forschungsgemeinschaft (DFG) and the
Stichting voor Fundamenteel Onderzoek der Materie (FOM),
financially supported by the Nederlandse Organisatie voor
Wetenschappelijk Onderzoek (NWO).
\end{theacknowledgments}

\bibliography{MuellerLudingPoeschel_AIP}

\begin{thebibliography}{36}
\expandafter\ifx\csname natexlab\endcsname\relax\def\natexlab#1{#1}\fi
\providecommand{\enquote}[1]{``#1''}
\expandafter\ifx\csname url\endcsname\relax
  \def\url#1{\texttt{#1}}\fi
\expandafter\ifx\csname urlprefix\endcsname\relax\def\urlprefix{URL }\fi
\providecommand{\eprint}[2][]{\url{#2}}

\bibitem[Thornton and Randall(1988)]{thornton88}
C.~Thornton, and C.~W. Randall, \enquote{Applications of theoretical contact
  mechanics to solid particle system simulation,} in \emph{Micromechanics of
  granular media}, Elsevier, Amsterdam, 1988.

\bibitem[Coppersmith et~al.(1996)]{coppersmith96}
S.~N. Coppersmith, C.~Liu, S.~Majumdar, O.~Narayan, and T.~A. Witten,
  \emph{Phys. Rev. E} \textbf{53}, 4673--4685 (1996).

\bibitem[Radjai et~al.(1996)]{radjai96b}
F.~Radjai, M.~Jean, J.~J. Moreau, and S.~Roux, \emph{Phys. Rev. Lett.}
  \textbf{77}, 274 (1996).

\bibitem[Radjai et~al.(1997)]{radjai97}
F.~Radjai, D.~Wolf, S.~Roux, M.~Jean, and J.~J. Moreau, \enquote{Force Networks
  in Granular Packings,} in \emph{Friction, Arching and Contact Dynamics},
  edited by D.~E. Wolf, and P.~Grassberger, World Scientific, Singapore, 1997.

\bibitem[Brockbank et~al.(1997)]{brockbank97}
R.~Brockbank, J.~M. Huntley, and R.~Ball, \emph{J. Phys. II France} \textbf{7},
  1521--1532 (1997).

\bibitem[Cates et~al.(1998)]{cates98b}
M.~E. Cates, J.~P. Wittmer, J.-P. Bouchaud, and P.~Claudin, \emph{Phys. Rev.
  Lett.} \textbf{81}, 1841--1844 (1998).

\bibitem[Radjai and Wolf(1998)]{radjai98}
F.~Radjai, and D.~E. Wolf, \emph{Granular Matter} \textbf{1}, 3--8 (1998).

\bibitem[Radjai et~al.(1998)]{radjai98b}
F.~Radjai, D.~E. Wolf, M.~Jean, and J.-J. Moreau, \emph{Phys. Rev. Lett.}
  \textbf{80}, 61--64 (1998).

\bibitem[Tsoungui et~al.(1998)]{tsoungui98b}
O.~Tsoungui, D.~Vallet, and J.-C. Charmet, \emph{Granular Matter} \textbf{1},
  65--69 (1998).

\bibitem[Tsoungui et~al.(1999)]{tsoungui99b}
O.~Tsoungui, D.~Vallet, and J.-C. Charmet, \emph{Phys. Rev. Lett.}  (1999).

\bibitem[Silbert et~al.(2001)]{silbert01}
L.~E. Silbert, D.~E. s, G.~S. Grest, T.~C. Halsey, and D.~Levine, \emph{Phys.
  Rev. E} \textbf{65}, 031304 (2001).

\bibitem[Blair et~al.(2001)]{blair01b}
D.~L. Blair, N.~W. Mueggenburg, A.~H. Marshall, H.~M. Jaeger, and S.~Nagel,
  \emph{Phys. Rev. E} \textbf{63}, 041304--1 (2001).

\bibitem[Snoeijer et~al.(2003)]{snoeijer03}
J.~H. Snoeijer, M.~van Hecke, E.~Somfai, and W.~van Saarloos, \emph{Phys. Rev.
  E} \textbf{67}, 030302(R) (2003).

\bibitem[Metzger(2004)]{metzger04}
P.~T. Metzger, \emph{Phys. Rev. E} \textbf{70}, 051303 (2004).

\bibitem[Snoeijer et~al.(2005)]{snoeijer05}
J.~H. Snoeijer, W.~G. Ellenbroek, T.~J.~H. Vlugt, and M.~van Hecke, Sheared
  force-networks: anisotropies, yielding and geometry (2005), cond-mat/0508411.

\bibitem[Liu et~al.(1995)]{liu95}
C.~Liu, S.~R. Nagel, D.~A. Schecter, S.~N. Coppersmith, S.~Majumdar,
  O.~Narayan, and T.~A. Witten, \emph{Science} \textbf{269}, 513 (1995).

\bibitem[Erikson et~al.(2002)]{erikson02}
J.~M. Erikson, N.~W. Mueggenburg, H.~M. Jaeger, and S.~R. Nagel, \emph{Phys.
  Rev. E} \textbf{66}, 040301(R) (2002).

\bibitem[Tighe et~al.(2008)]{tighe08}
B.~P. Tighe, A.~R.~T. van Eerd, and T.~J.~H. Vlugt, \emph{Phys. Rev. Lett.}
  \textbf{100}, 238001 (2008).

\bibitem[Donev et~al.(2004)]{donev04}
A.~Donev, I.~Cisse, D.~Sachs, E.~Variano, F.~H. Stillinger, R.~Connelly,
  S.~Torquato, and P.~M. Chaikin, \emph{Science} \textbf{303}, 990--993 (2004).

\bibitem[Corwin et~al.(2005)]{corwin05}
E.~I. Corwin, H.~M. Jaeger, and S.~R. Nagel, \emph{Nature} \textbf{435},
  1075--1078 (2005).

\bibitem[Dauchot and Marty(2005)]{dauchot05}
O.~Dauchot, and G.~Marty, \emph{Phys. Rev. Lett.} \textbf{95}, 265701 (2005).

\bibitem[Song et~al.(2005)]{song05}
C.~Song, P.~Wang, F.~Potiguar, and H.~Makse, \emph{J. Phys.: Condens. Matter}
  \textbf{17}, S2755--S2770 (2005).

\bibitem[Silbert et~al.(2005)]{silbert05}
L.~Silbert, A.~J. Liu, and S.~R. Nagel, \emph{Phys. Rev. Lett.} \textbf{95},
  098301 (2005).

\bibitem[Majmudar et~al.(2007)]{majmudar07}
T.~S. Majmudar, M.~Sperl, S.~Luding, and R.~P. Behringer, \emph{Phys. Rev.
  Lett.} \textbf{98}, 058001 (2007).

\bibitem[Aranson et~al.(2008)]{aranson08}
I.~S. Aranson, L.~S. Tsimring, F.~Malloggi, and E.~Cl\'ement, \emph{Phys. Rev.
  E} \textbf{78}, 031303 (2008).

\bibitem[Zeravcic et~al.(2009)]{zeravcic09}
Z.~Zeravcic, N.~Xu, A.~J. Liu, S.~R. Nagel, and W.~van Saarloos,
  \emph{Europhys. Lett.} \textbf{87}, 26001 (2009).

\bibitem[Mailman et~al.(2009)]{mailman09}
M.~Mailman, C.~F. Schreck, C.~S. O'Hern, and B.~Chakraborty, \emph{Phys. Rev.
  Lett.} \textbf{102}, 255501 (2009).

\bibitem[Maloney(2006)]{maloney06}
C.~E. Maloney, \emph{Phys. Rev. Lett.} \textbf{97}, 035503 (2006).

\bibitem[Tanguy et~al.(2002)]{tanguy02}
A.~Tanguy, J.~Wittmer, F.~Leonforte, and J.-L. Barrat, \emph{Physical Review B}
  \textbf{66}, 174205--1--174205--17 (2002).

\bibitem[Silbert and Silbert(2009)]{silbert09}
L.~E. Silbert, and M.~Silbert, \emph{Phys. Rev. E} \textbf{80}, 041304 (2009).

\bibitem[Jakeman(1980)]{jakeman80}
E.~Jakeman, \emph{J. Phys. A: Math. Gen.} \textbf{13}, 31--48 (1980).

\bibitem[Mueth et~al.(1998)]{mueth98}
D.~M. Mueth, H.~M. Jaeger, and S.~R. Nagel, \emph{Phys. Rev. E.} \textbf{57},
  3164--3169 (1998).

\bibitem[Blair and Kudrolli(2001)]{blair01}
D.~L. Blair, and A.~Kudrolli, \emph{Phys. Rev. E} \textbf{64}, 050301 (2001).

\bibitem[Allen and Tildesley(1987)]{allen87}
M.~P. Allen, and D.~J. Tildesley, \emph{Computer Simulation of Liquids}, Oxford
  University Press, Oxford, 1987.

\bibitem[Luding(1998)]{luding98c}
S.~Luding, \enquote{Collisions \& Contacts between two particles,} in
  \emph{Physics of dry granular media - NATO ASI Series E350}, edited by H.~J.
  Herrmann, J.-P. Hovi, and S.~Luding, Kluwer Academic Publishers, Dordrecht,
  1998, p. 285.

\bibitem[Bagi(2003)]{bagi03}
K.~Bagi, \emph{Granular Matter} \textbf{5}, 45--54 (2003).

\end{thebibliography}

\bibliographystyle{aipproc}   

\IfFileExists{\jobname.bbl}{}
 {\typeout{}
  \typeout{******************************************}
  \typeout{** Please run "bibtex \jobname" to optain}
  \typeout{** the bibliography and then re-run LaTeX}
  \typeout{** twice to fix the references!}
  \typeout{******************************************}
  \typeout{}
 }

\end{document}